\documentclass[preprintnumbers,prd,twocolumn,showpacs,floatfix,nofootinbib,superscriptaddress]{revtex4}
\usepackage{graphicx}
\usepackage{epsfig}
\usepackage{bm}
\usepackage{amssymb}
\usepackage{float}
\usepackage{amsmath}
\usepackage{dcolumn}
\usepackage{cancel}
\usepackage[colorlinks]{hyperref}
\usepackage[usenames,dvipsnames]{color}
\hypersetup{
     breaklinks=true,
    pdfstartview={FitH},    % fits the width of the page to the window
    colorlinks=true,       % false: boxed links; true: colored links
    linkcolor=blue,          % color of internal links
    citecolor=red,        % color of links to bibliography
    filecolor=magenta,      % color of file links
    urlcolor=blue,           % color of external links
    anchorcolor=green,      % Color for anchor text
    linktocpage=true
}

\begin{document}
\title{
A note on superposition of two unknown states using Deutsch CTC model
%A note on superposition of two unknown states in Deutsch CTC model
}% Force line breaks with \\
\author{Sasha Sami}
\affiliation{  Center for Computational Natural Sciences and Bio-informatics, IIIT-Hyderabad.}
\author{Indranil Chakrabarty }
\affiliation{ Center for Security Theory and Algorithmic Research, IIIT-Hyderabad.}

\date{\today}% It is always \today, today,
\begin{abstract}
\noindent In a recent work, authors prove a yet another no-go theorem that
forbids the existence of a universal probabilistic quantum protocol producing a superposition of two unknown quantum states. In this short note, we show that in the presence of closed time like curves, one can indeed create superposition of unknown quantum states and evade the no-go result.
%We prove this within the quantum mechanical framework in the Deutsch CTC model (D-CTC).
%We also prove that in a device independent framework if two inputs are unknown quantum states, then by applying the CTC condition one can obtain superposition of inputs in the causality respecting region.    
\end{abstract}

\pacs{Valid PACS appear here}% PACS, the Physics and Astronomy
\maketitle

%\tableofcontents

\section{Introduction}
\noindent In the past two decades the quantum information theory played an important role in achieving a huge range of information processing tasks which are still impossible to achieve with the current set of technologies available in the classical world \cite{ball1,ball2,ball3,ball4,ball5,ball6,ball7,ball8,ball9,ball10,ball11,ball12,ball13,ball14,ball15,ball16,ball17,ball18}. At the same time there are certain tasks in the classical world which are impossible to execute with the quantum resources. These impossible operations are termed as no-go theorems of quantum information theory \cite{all1,all2,all3,all4,all5,all6,all7,all8,all9} and indeed they play a very crucial role in the security and privacy aspect of the quantum technology. 
A good example in this context is the no-cloning theorem , which states that non-orthogonal quantum states cannot be cloned which serves as an underlying reason 
for the existence of secure quantum cryptography. \textcolor{black} { Quantum information provides different computational resource than classical information. In a pioneering work by Alvarez-Rodriguez et al \cite{Alv}, it was demonstrated that there is no unitary protocol which would allow to add quantum states belonging to different Hilbert space. This analysis was further extended recently where researchers articulate the fact that in spite superposition being an intriguing phenomenon of quantum physics, it is impossible to create an arbitrary superposition of unknown quantum states\cite{Osz}. Circumventing the said no-go is a challenging problem and is the main motivation of the present letter.  } \\

\noindent General theory of relativity does allow the existence of closed
timelike curves (CTCs), which is 
a world line that connects back to itself \cite{stockum, godel, deser, ori, mor,fro,kim,got,haw, politzer} In other words in  the  presence
of a space time wormhole these word lines could link a future space time
point with a past world point. 
The latter would give rise to chronological paradoxes, for instance the “grandfather paradox”. The important question is whether we can have a computationally efficient model of CTC where these paradoxes are resolved. As  an immediate answer to this question Deutsch proposed a computational model of
quantum systems in the presence of CTCs. These paradoxes are resolved by presenting a method for finding self-consistent solutions of CTC  interactions in quantum theory  \cite{deu} (cf. \cite{pctc1,pctc2,pctc3,pctc4,pctc5,pctc6,pctc7}).\\   
%(in the presence
%of a space time wormhole).
% that could link a future space time point
%with a past space time point. 
%Classically even though such exotic objects are allowed in physics they lead to 
%various paradoxes. But we know that at a fundamental level quantum theory is 
%the correct theory. % His formulation can 
% resolve various classical paradoxes and provides a self-consistent solutions of
% CTC interactions in quantum theory. 

\noindent 
Further investigations revealed that the presence of closed time like curves can significantly affect the computational and other abilities of a system
\cite{bru,bac,aar}. 
These include factorization of composite numbers
efficiently with the help of a classical computer \cite{bru} and ability to solve NP-complete problems \cite{bac}.
%- an impossibility for a quantum computer.\\ 
%Another important finding in this direction is that computational complexity is that the power of a
%polynomial time bounded computer in classical or quantum world 
%assisted by CTCs can be solved in a space polynomial in the problem size 
%(PSPACE) but potentially exponential time \cite{aar}.
%
Brun {\em et al.} \cite{bru1} have shown that with the access to CTCs, it is possible to perfectly distinguish non orthogonal quantum states, having wide range of implications for the security of quantum cryptography. In another work \cite{ralph}, the information flow of quantum states interacting with closed time like curves was investigated.
% %a better picture of Deutsch CTC model was given by  
% tracking of the information flow of quantum system at the time of its interaction with CTC's.
% Their work also supports
% the concept that CTC assisted party have the power of distinguishing non-orthogonal quantum 
% states \cite{bru1}.
%However, r
Few years back, it has been shown that the presence of CTC 
has implications for purification of mixed states \cite{akp}, and in making non local no signaling boxes  to signaling boxes \cite{indra}. Recently. it was demonstrated  that teleportation of quantum information, even in its approximate version, from a CR region to a CTC region is disallowed \cite{asu}. 
In a paper \cite{ben1},  Bennett {\it et al.}  have argued against the Deutsch model (D-CTC) and opined for revisiting the  implications obtained by assuming the existence of CTCs as described by Deutsch model. \\

%\noindent \textit{~~~~~~~~~~~~~~~~~~~A. Deutsch CTC: }\\

\noindent Qubits having a closed time like world line   can give rise to various paradoxes. A predominant one of them is the grandfather paradox. However these paradoxes can be avoided by using the self consistency condition of the D-CTC model.
The Deutsch self-consistency conditions have two components to it: one qubit from the chronology respecting region (CR) and another qubit having a word line like a closed time like curve which we will refer as CTC qubit. This condition demands the initial density matrix of a CTC system must be equal to its output density matrix after it has interacted with a chronology respecting system CR under a unitary operation $U$,

\begin{equation}
\label{eq:Deutsch_1}
\rho_{CTC}= Tr_{CR}\{U(\rho_{CR} \otimes \rho_{CTC} )U^{\dagger}\}, 
\end{equation}
\begin{equation}
\label{eq:Deutsch_2}
\rho_{out}= Tr_{CTC}\{U(\rho_{CR} \otimes \rho_{CTC} )U^{\dagger}\}.
\end{equation}
In Eq. (\ref{eq:Deutsch_1}) $\rho_{CTC}$ stands for the density matrix of the CTC system before interaction and the right hand side
of the equation gives the partial density matrix of the CTC system after interaction. In Eq.\ref{eq:Deutsch_2} $\rho_{out}$ gives the density matrix of the chronology respecting system (CR) after interaction, whose  initial density matrix  $\rho_{CR}$. \textcolor{black}{Let us reiterate that though CTC's are problematic in Einstein's theory of relativity due to violation of principle of causality, the problem is circumvented in quantum framework. As demonstrated in \cite{cas,al,rin,zhang}  the affect of CTC's can be mimicked using classical and quantum simulation. For instance in Ref.\cite{rin}, the authors
succeeded in 
experimentally simulating the nonlinear behaviour of a qubit interacting in a unitary fashion with an older
version of itself. The presence of CTCs, in particular, allowed them to discriminate the non-orthogonal states which is otherwise impossible. They also examined the other no-go results leaving aside the problem of superposition of unknown quantum states.
} 

\noindent In this work we show that if we have access to a closed time like curve satisfying Deutsch kinematic conditions then we can indeed design an unitary operator which will be able to create a superposition of two unknown quantum states.

\noindent According to a recent no go theorem \textcolor{black}{\cite{Alv,Osz}},  given two unknown quantum states $|\phi_1\rangle \langle \phi_1|$ and $|\phi_2\rangle \langle \phi_2|$ it is not possible to create the state $|\phi \rangle\langle \phi|$ where $|\phi\rangle = \gamma^{-1} ( \alpha |\phi_1\rangle + \beta |\phi_2\rangle)$, where $\gamma$ is the normalizing factor and $\alpha,\beta$ are given complex numbers. A probabilistic protocol is also given, to create superposition of two unknown states where the class of input states for which superposition is to be created is given along with information from which superposition has to be generated. But with the assistance of Deutsch CTC we can create superposition of two unknown states deterministically, corresponding to fixed complex numbers $\alpha$ and $\beta$, if the set $\{\psi_j\}$ from which the two unknown states are taken is known before hand. The method follows directly from the proof of distinguishability  of non-orthogonal  states with under Deutsch CTC.
%\noindent \subsection{Quantum Mechanical Framework}
As shown by  Brun \emph{et al.}\cite{bru1} if $\{|\psi_j\rangle \}_{j=0}^{N-1}$ is a set of $N$ distinct states in a $N$ dimensional space, then by using Deutsch CTC we can implement the mapping $\forall j \; |\psi_j\rangle \to |j\rangle$, where $|j\rangle$ forms an orthonormal basis for the $N$ dimensional space. The unitary operation they used to carry out this transformation is a SWAP operation
followed by a controlled unitary operation from chronology respecting system to the CTC system given by,
\begin{equation}
\label{Brun_big_unitary}
U=\sum_{k=0}^{N-1}|k\rangle \langle k| \otimes U_k,
\end{equation}
where $U_k$ are unitary operations that satisfy the following conditions: $(1)$ $U_k |\psi_k\rangle = |k\rangle$ for $0\le k < N$ and $(2)$ $\langle j |U_k|\psi_j\rangle \ne 0$ for  $0 \le j,k < N$. The latter conditions come from constraint of unique solution to the Deutsch self-consistency condition. Brun \emph{et al.} showed that it is always possible to construct unitary operations $U_k$ satisfying constraints $(1)$ and $(2)$ and gave a method for the same. It can be checked that if initially the chronologically respecting system is in state $|\psi_j\rangle$ then after interacting it with the CTC system under the unitary operation given by Eq. (\ref{Brun_big_unitary}) the final state of both CR and CTC system is 
\begin{eqnarray}
&&\rho_{out}=\rho_{CTC}=|j\rangle \langle j|
\nonumber \\
&&(|\psi_j\rangle\langle \psi_j|\otimes\rho_{CTC}= |\psi_j\rangle\langle \psi_j|\otimes |j\rangle\langle j| \nonumber \\
&&\rightarrow_{(SWAP)}\rightarrow  |j\rangle\langle j|\otimes |\psi_j\rangle\langle \psi_j|\rightarrow_{(U)}\rightarrow |j\rangle\langle j|\otimes |j\rangle\langle j|) \;\;\;\;\;\;
\end{eqnarray}
Here we also follow the  similar setup.  Let $|\phi_1\rangle=|\psi_m\rangle $ and $|\phi_2\rangle=|\psi_n\rangle$ be two unknown states from the set $\{|\psi_j\rangle \}_{j=0}^{N-1}$ for which we wish to create the superimposition $|\phi\rangle = \gamma^{-1} ( \alpha |\phi_1\rangle + \beta |\phi_2\rangle)$ where $\gamma$ is the normalizing factor and $\alpha,\beta$ are given complex numbers. For this we require two CTC systems, for each of these states $|\phi_1\rangle$ and $|\phi_2\rangle$. To do so, we interact both the states with separate CTC systems under unitary given by Eq.\ref{Brun_big_unitary}. Let the states of chronological respecting systems in both the cases after interaction with their respective CTC systems be $\rho_{out_1}=|m\rangle \langle m|$ and $\rho_{out_2}=|n\rangle \langle n|$ respectively. Let $U^{'}$ be a unitary defined as

\begin{equation}
\label{eq:unitary}
U^{'} = \sum _{i,j=0}^{N-1} |i\rangle \langle i| \otimes |j \rangle \langle j | \otimes U_{\alpha,\beta}^{i,j}
\end{equation}
where  $U_{\alpha,\beta}^{i,j}$ are unitary operations for  $0 \le i,j < N$, such that,
\begin{eqnarray}
 U_{\alpha,\beta}^{i,j}|0\rangle =|\omega\rangle_{\alpha,\beta}^{i,j}=\gamma^{-1}(\alpha |\psi_i\rangle + \beta |\psi_j\rangle) 
\end{eqnarray}
for some fixed state $|0\rangle$. Such unitary operations  $U_{\alpha,\beta}^{i,j}$ can always be constructed by Gram Schmidt process on the set $S=|\omega\rangle_{\alpha,\beta}^{i,j} \cup \{|\psi_j\rangle \}_{j=0}^{N-1} $ with the first element for the process being $|\omega\rangle_{\alpha,\beta}^{i,j}$. If $S$ does not contain $N$ linearly independent states, the orthonormal states obtained by the process can always be extended.  If the input states are the same that is $|\phi_1\rangle=|\phi_2\rangle$ then the desired superposition is same as the input states. So for simplicity $U_{\alpha,\beta}^{i,i}={U_i}^{-1}P_i$ where $P_i$ is a permutation unitary such that $P_i|0\rangle=|i\rangle$ and ${U_i}^{-1}$ is the inverse of the
unitary $U_i$ given by $U_k$ in Eq (\ref{Brun_big_unitary}). When the unitary $U^{'}$ defined by Eq (\ref{eq:unitary}) is applied on $\rho_{out_1} \otimes \rho_{out_2} \otimes |0\rangle \langle 0| $ ( where $|0\rangle $ is the fixed ancilla state defined above ) then the desired superimposition of $|\phi_1\rangle$ and $|\phi_2\rangle$ for the given complex numbers $\alpha,\beta$ is obtained on the ancilla system. 
\begin{equation} \label{proof}
\begin{split}
U^{'}(\rho_{out_1} \otimes \rho_{out_2} \otimes |0\rangle \langle 0|) & = U^{'}(|m\rangle \langle m| \otimes |n\rangle \langle n| \otimes |0\rangle \langle 0|) \\
 & = |m\rangle \langle m| \otimes |n\rangle \langle n| \otimes U_{\alpha,\beta}^{i,j}|0\rangle \langle 0| \\
 & = |m\rangle \langle m| \otimes |n\rangle \langle n| \otimes | \omega_{\alpha,\beta}^{i,j}\rangle \langle \omega_{\alpha,\beta}^{i,j}|
\end{split}
\end{equation}

\noindent \textit{Example:} Now consider an example where $N=2$ and the given set of
distinct states is $\{|0\rangle,|-\rangle\}$. And let $\alpha,\beta$ be the given complex numbers. In this case the unitary given by Eq. (\ref{Brun_big_unitary}) reduces to $U=|0\rangle \langle0|\otimes I + |1\rangle \langle1|\otimes H$ where $I$ and $H$ are identity and Hadamard operators. And the unitary operators $U_{\alpha,\beta}^{i,j}$ reduce to 
%\begin{eqnarray}
%$U_{\alpha,\beta}^{0,0}=I$, $U_{\alpha,\beta}^{0,1}=\frac{1}{\gamma_1}\begin{bmatrix}
 %       \alpha + \frac{\beta}{\sqrt{2}} & \frac{\beta^{*}}{\sqrt{2}} \\
  %    -\frac{\beta}{\sqrt{2}}  & \alpha^{*} + \frac{\beta^{*}}{\sqrt{2}}                     
   %  \end{bmatrix}$
    % $U_{\alpha,\beta}^{1,0}=\frac{1}{\gamma_2}\begin{bmatrix}
     %   \beta + \frac{\alpha}{\sqrt{2}} & \frac{\alpha^{*}}{\sqrt{2}} \\
      %-\frac{\alpha}{\sqrt{2}}  & \beta^{*} + \frac{\alpha^{*}}{\sqrt{2}}                     
     %\end{bmatrix}$ and $U_{\alpha,\beta}^{1,1}=HX$,
%\end{eqnarray}     
 %     where $\gamma_1={((\alpha+\frac{\beta}{\sqrt{2}})^2+\frac{\beta^2}{2}})^{\frac{1}{2}}$ and $\gamma_2={((\beta+\frac{\alpha}{\sqrt{2}})^2+\frac{\alpha^2}{2}})^{\frac{1}{2}}$ are normalizing factors and $X$ is the phase flip operator. Using Eq's. (\ref{eq:unitary}) and (\ref{proof}), it can be checked that for values of $i,j$ the desired superposition is created.\\
  %%%%%%
 \begin{eqnarray}
&& U_{\alpha,\beta}^{0,0}=I\\ 
&& U_{\alpha,\beta}^{0,1}=\frac{1}{\gamma_1}\begin{bmatrix}
        \alpha + \frac{\beta}{\sqrt{2}} & \frac{\beta^{*}}{\sqrt{2}} \\
      -\frac{\beta}{\sqrt{2}}  & \alpha^{*} + \frac{\beta^{*}}{\sqrt{2}}                     
     \end{bmatrix}\\
    && U_{\alpha,\beta}^{1,0}=\frac{1}{\gamma_2}\begin{bmatrix}
        \beta + \frac{\alpha}{\sqrt{2}} & \frac{\alpha^{*}}{\sqrt{2}} \\
      -\frac{\alpha}{\sqrt{2}}  & \beta^{*} + \frac{\alpha^{*}}{\sqrt{2}}                     
     \end{bmatrix} \\
 &&    U_{\alpha,\beta}^{1,1}=HX, 
     \end{eqnarray}
     where $\gamma_1={((\alpha+\frac{\beta}{\sqrt{2}})^2+\frac{\beta^2}{2}})^{\frac{1}{2}}$ and $\gamma_2={((\beta+\frac{\alpha}{\sqrt{2}})^2+\frac{\alpha^2}{2}})^{\frac{1}{2}}$ are normalizing factors and $X$ is the phase flip operator. Using Eq's. (\ref{eq:unitary}) and (\ref{proof}), it can be checked that for values of $i,j$ the desired superposition is created.

  %%%%%%%
\noindent In this letter,  we have shown that creating superposition of an unknown state is possible in causality respecting region provided we allow the interaction with a closed time like curve.  This once again shows the enormous power of closed time like curves in making things possible which are otherwise impossible in the chronology respecting regions. \color{black}{It would be  interesting to see how by mimicking the effect of CTC by classical and quantum simulation \cite{rin} we can create a superposition of two unknown quantum states. Simultaneously it will be fascinating to see whether in such a situation if we are not able to get perfect superpostion, then can we get a fidelity more than the fidelity obtained in case of an approximate quantum adder\cite{Alv}.  Both of these questions are very interesting and we defer it to our future investigations}.   
\section{Acknowledgment}\noindent  We are indebted to A. K. Pati for suggesting the problem and for his constant help and encouragement thereafter.
     
%\noindent \subsection{Device Independent Framework}

%\section{Conclusions}

\end{document}